# First Experimental Demonstration of Gate-all-around III-V MOSFETs by Top-down Approach

J. J. Gu [1)], Y. Q. Liu [2)], Y. Q. Wu [1)], R. Colby [3)], R. G. Gordon [2)], and P. D. Ye [1)]

[1)] *School of Electrical and Computer Engineering, Purdue University, West Lafayette, IN 47906, U.S.A.*
[2)] *Department of Chemistry and Chemical Biology, Harvard University, Cambridge, MA 02138, U.S.A.*
[3)] *School of Materials Engineering, Purdue University, West Lafayette, IN 47907, U.S.A.*
Tel: 1-765-494-7611, Fax: 1-765-496-7443, Email: yep@purdue.edu, jjgu@purdue.edu

## Abstract

The first inversion-mode gate-all-around (GAA) III-V MOSFETs are experimentally demonstrated with a high mobility $In_{0.53}Ga_{0.47}As$ channel and atomic-layer-deposited (ALD) $Al_2O_3$/WN gate stacks by a top-down approach. A well-controlled InGaAs nanowire release process and a novel ALD high-*k*/metal gate process has been developed to enable the fabrication of III-V GAA MOSFETs. Well-behaved on-state and off-state performance has been achieved with channel length ($L_{ch}$) down to 50nm. A detailed scaling metrics study (*S.S.*, *DIBL*, $V_T$) with $L_{ch}$ of 50nm - 110nm and fin width ($W_{Fin}$) of 30nm - 50nm are carried out, showing the immunity to short channel effects with the advanced 3D structure. The GAA structure has provided a viable path towards ultimate scaling of III-V MOSFETs.

## Introduction

Recently, continuous progress has been made in the understanding and improvement of high-k/III-V interfaces. However, to realize III-V FETs beyond the 14nm technology node, emerging 3D device structures are necessary to suppress short-channel effects (SCE). III-V FinFETs [1-2] as well as multi-gate quantum-well FETs [3] have been shown to improve greatly the off-state performance of III-V FETs with deep submicron gate lengths. On the other hand, the GAA structure has been proven [4-6] on Si CMOS to be the most resistant to SCE, thanks to having the best gate electrostatic control. Therefore, a III-V GAA FET is the most promising candidate for the ultimately scaled III-V FETs. In this paper, we report *the first experimental demonstration of inversion-mode* $In_{0.53}Ga_{0.47}As$ *GAA FETs by a top-down approach* with ALD $Al_2O_3$/WN gate stacks. Benefiting from the GAA structure, we have demonstrated the shortest channel length ($L_{ch}$ = 50nm) III-V MOSFETs to date with well-behaved on- and off-state characteristics. A systematic scaling metrics study has been carried out for $In_{0.53}Ga_{0.47}As$ GAA FETs with $L_{ch}$ from 110nm down to 50nm, $W_{Fin}$ of 30nm and 50nm, fin height ($H_{Fin}$) of 30nm, and wire lengths ($L_{NW}$) of 150 to 200nm. The SCE has been effectively suppressed by the advanced 3D design.

## Experiment

Fig. 1 shows a schematic view of an $In_{0.53}Ga_{0.47}As$ GAA FET fabricated in this work. Table 1 and Fig. 2 depict the key fabrication processes for $In_{0.53}Ga_{0.47}As$ GAA FETs. A 30nm p-doped $2\times10^{16}$ $cm^{-3}$ $In_{0.53}Ga_{0.47}As$ channel layer was epitaxially grown on a p+ (100) InP substrate by MBE as the starting material (Fig. 2-1). After surface degrease and $NH_4OH$ pretreatment, 10nm $Al_2O_3$ was grown by ALD as an encapsulation layer. Source/drain Si implantation was then performed at an energy of 20keV and a dose of $1\times10^{14}$ $cm^{-2}$ (Fig. 2-2). The dopant activation was carried out at 600 °C for 15 seconds in nitrogen ambient. The source/drain separation determines the final $L_{ch}$ of the devices. After removing the encapsulation layer by buffered oxide etch (BOE), the InGaAs fin etching was done by $BCl_3$/Ar high density plasma etching (HDPE) [1] (Fig. 2-3). The diluted ZEP520A electron-beam resist with a thickness of 100 nm was used as a hard mask for the fin etching and the smallest $W_{Fin}$ defined was 30nm. After surface cleaning in BOE and diluted $HCl:H_2O_2$ solution, the InGaAs channel release process was carried out using $HCl:H_2O$ (1:2) solution (Fig. 2-4). HCl based solution can selectively etch InP over InGaAs. However, the etching is found to be highly anisotropic. Therefore the InGaAs fins have to be patterned along <100> directions for a successful release process [7]. Fig. 3 (a) shows the tilted SEM image of free-standing InGaAs nanowires after channel release process. Fig. 3 (b) shows the cross-sectional STEM image of InGaAs nanowire test structures wrapped by 50nm ALD $Al_2O_3$ on InP substrate, confirming the nanowires are completely released. Fig. 4 shows the different fin patterning direction and the corresponding etching profile. Undercut etching is achieved with fin patterned along [010] direction. This determined the final device alignment to the substrate and the transport direction of devices (along [010]). After channel release, the samples were soaked in 20% $(NH_4)_2S$ for pre-gate interface passivation. Then the samples were immediately transferred to an ASM F-120 ALD reactor via room ambient. 10nm $Al_2O_3$ was regrown as the gate dielectric at 300 °C. 20nm WN metal gate was then deposited in a separate ALD reactor at 385 °C (Fig. 2-5), with a resistivity of ~4000μΩ·cm [8]. The conformal deposition of ALD $Al_2O_3$/WN surrounding the nanowire channel is the key fabrication process for realizing the GAA structure. After gate stack deposition, gate etch process was performed using $CF_4$/Ar HDPE, where Cr/Au gate pattern was defined as the hard mask (Fig. 2-6). The $CF_4$ based dry etching chemistry provides excellent selectivity between WN and $Al_2O_3$, resulting in a damage-free gate oxide. The source/drain contact was then formed by electron beam evaporation of Au/Ge/Ni, followed by 350 °C rapid thermal annealing in nitrogen ambient. Finally, the Ti/Au test pads were defined. The fabricated MOSFETs have a nominal $L_{ch}$ varying from 50nm to 120nm, $W_{Fin}$ from 30nm to 50nm, and different numbers of parallel channels (1 wire, 4 wires, 9 wires or 19 wires). Fig. 3 (c) shows the SEM image of a InGaAs GAA FET with 4 parallel wires. All patterns were defined by a Vistec VB-6 UHR electron-beam lithography system. A Keithley 4200 was used to measure MOSFET output characteristics.

## Results and Discussion

Fig. 5 - 6 show the well-behaved output and transfer characteristics as well as $I_g$-$V_g$ of a $L_{ch}$ = 50nm GAA FET. The current here is normalized by the total perimeter of the $In_{0.53}Ga_{0.47}As$ channel, i.e. $W_G = (2W_{Fin} + 2H_{Fin}) \times$ (No. of wires). A representative 50nm $L_{ch}$ device shows on-current of 720μA/μm, transconductance of 510μS/μm and reasonable off-state characteristics with subthreshold swing (SS) of 150mV/dec and drain-induced barrier lowering (DIBL) of 210mV/V. Although operating in inversion-mode, the threshold voltage of the device is -0.68V from linear extrapolation at $V_{ds}$=50mV due to the relatively low work function of ALD WN metal (~4.6eV). Due to the junction leakage current and a very large area ratio (>$10^3$) between implanted junction and GAA channels, the source current is used to obtain the intrinsic current in the channel. Gate leakage current is minimal in the entire gate voltage range, indicating 10nm $Al_2O_3$ is sufficient for GAA structure and further equivalent oxide thickness (EOT) scaling is achievable. It also shows that the WN gate etch process is damage-free. Fig. 7 shows the extrinsic and intrinsic transconductance at $V_{ds}$ = 1V for the same device. The source/drain resistance $R_{SD}$ is extracted to be around 1150Ω·μm. The maximum intrinsic transconductance is 750μS/μm. The relatively large $R_{SD}$ results from the non-optimal implantation process, the additional extension resistance at the nanowire channel / source drain link region, and the non-self-aligned process. The EOT of the device is estimated to be 4.5nm. Further reduction of the EOT can be achieved by reducing the oxide thickness or integrate higher-*k* dielectrics. A simple calculation reveals that by reducing the ALD $Al_2O_3$ thickness to 5nm or 2.5nm [9], the SS of the 50nm $L_{ch}$ device can be improved to 105mV/dec or 82mV/dec, respectively. Therefore a SS lower than 100mV/dec is achievable on surface-channel InGaAs MOSFETs at sub-100nm channel lengths with the current $(NH_4)_2S$ surface passivation technique.

Fig. 8 shows the $I_{ON}$ and $g_m$ scaling metrics for $L_{ch}$ = 50 - 110nm and $W_{Fin}$ = 30nm. The values are determined by measuring 20 different devices at the same $L_{ch}$. Error bar shows the statistical variation of multiple devices. The $I_{ON}$ and $g_m$ steadily increase with scaling of $L_{ch}$. A low field mobility ($\mu_0$) of 903cm$^2$/V·s is extracted using Y-function method for the 110nm $L_{ch}$ device, showing 3× mobility enhancement compared to bulk Si NMOSFETs at similar $L_{ch}$ [10]. Fig. 9 - 11 show the off-state ($V_T$, SS and DIBL) scaling metrics for $L_{ch}$ = 50 - 110nm with $W_{Fin}$=30nm and 50nm. $V_T$ is determined using 1μA/μm metrics at $V_{ds}$ = 50mV. From Fig. 9, 30nm $W_{Fin}$ devices show better $V_T$ roll-off properties when $L_{ch}$ is shrinking. The SS for 30nm $W_{Fin}$ devices are almost unchanged at around 150mV/dec when scaling $L_{ch}$ down to 50nm, indicating excellent control of SCE, whereas the 50nm $W_{Fin}$ devices show larger SS, which increases with scaling of $L_{ch}$. It is noted here that the 100nm $L_{ch}$ InGaAs FinFET with 5nm $Al_2O_3$ gate oxide shows similar SS [1] as the 50nm $L_{ch}$ GAA FET with 10nm $Al_2O_3$ in this work. This translates to at least a factor of 2 improvement of midgap $D_{it}$ (~$5.6\times10^{12}$/cm$^2$·eV) achieved. The improved interface quality indicates that the newly-developed channel release process can provide a smooth damage-free InGaAs bottom surface. Fig. 11 shows that 30nm $W_{Fin}$ devices have smaller DIBL and the DIBL is roughly independent of $L_{ch}$, confirming the effective SCE control. Further SS and DIBL reduction can be achieved by scaling down EOT and reducing the InGaAs nanowire dimension. Fig. 12 shows the transfer characteristics for two GAA FETs with 1 wire and 4 wires in parallel, respectively. Fig. 13 – 14 show a linear relationship of $I_{s,max}$ and $g_{m,max}$ with the number of wires in both the linear and saturation regimes. The linear dependence demonstrates that the fabrication process is scalable towards high integration density. Each wire can deliver an $I_{sat}$ = 90μA and $g_m$ = 66μS at $V_{ds}$=1V. Fig. 15 shows the output characteristic for a *hero* GAA FET with $I_{ON}$ = 1.17mA/μm and $g_{m,max}$ = 701μS/μm. An $I_{on}$ over 4.5mA/μm can be obtained if the current is normalized by $W_{Fin}$.

Fig. 16 benchmarks the $g_m$·EOT product vs. $L_{ch}$ of $In_{0.53}Ga_{0.47}As$ GAA FETs with our previous work on InGaAs MOSFETs [1][9][11-13]. Despite the low indium concentration (53%), the $In_{0.53}Ga_{0.47}As$ GAA MOSFETs show the highest $g_m$·EOT product. Better on-state performance is expected on InGaAs GAA MOSFETs with higher indium content (65% to 75%), due to the higher electron mobility and the charge neutral level being closer to the conduction band edge. Table 2 compares the device structure and performance of $In_{0.53}Ga_{0.47}As$ GAA FETs in this work with representative top-down non-planar III-V FETs [1-3][11]. Due to the excellent electrostatic control of the channel by GAA structure, $L_{ch}$ has been pushed down to 50nm with excellent on-state and reasonable off-state performance.

## Conclusion

We have demonstrated for the first time inversion-mode $In_{0.53}Ga_{0.47}As$ GAA MOSFETs with ALD $Al_2O_3$/WN gate stacks. The highest saturation current reaches 1.17mA/μm at $L_{ch}$ = 50nm and $V_{ds}$ = 1V with $g_{m,max}$ = 701μS/μm. Detailed scaling metrics study shows that the 3D GAA structure can effectively control the SCE with $L_{ch}$ scaling down to at least 50nm, making III-V GAA FET a very promising candidate for ultimately scaled III-V logic device technology.

## Acknowledgement

The authors would like to thank R. Wang, A. T. Neal, M. Xu, M. Luisier, M. S. Lundstrom, D. A. Antoniadis, J. del Alamo, C. Zhang, and X. L. Li for valuable discussions and technical assistances. This work is supported by the SRC FCRP MSD Focus Center and US National Science Foundation.

## References

[1] Y. Q. Wu et al., *IEDM Tech. Dig.* 331 (2009).
[2] H. -C. Chin et al., *IEEE Electron Device Lett.* **32**, 146 (2011).
[3] M. Radosavljevic et al., *IEDM Tech. Dig.* 126 (2010).
[4] Y. Tian et al., *IEDM Tech. Dig.* 895 (2007).
[5] N. Singh et al., *IEEE Trans. Electron Device*, **55**, 3107 (2008).
[6] S. D. Suk et al., *IEDM Tech. Dig.* 552 (2005).
[7] J. J. Gu et al., *Applied Physics Letters*, in press.
[8] J. S. Becker et. al., *Chem. Mat.,* **15**, 2969 (2003).
[9] Y. Q. Wu et al., *IEDM Tech. Dig.* 323 (2009).
[10] A. Cros et al., *IEDM Tech. Dig.* 663 (2006)
[11] Y. Xuan et al., *IEDM Tech. Dig.* 637 (2007).
[12] Y. Xuan et al., *IEEE Electron Device Lett.,* **29**, 294 (2008).
[13] J. J. Gu et al., *J. of Appl. Phys.* **109**, 053709 (2011).

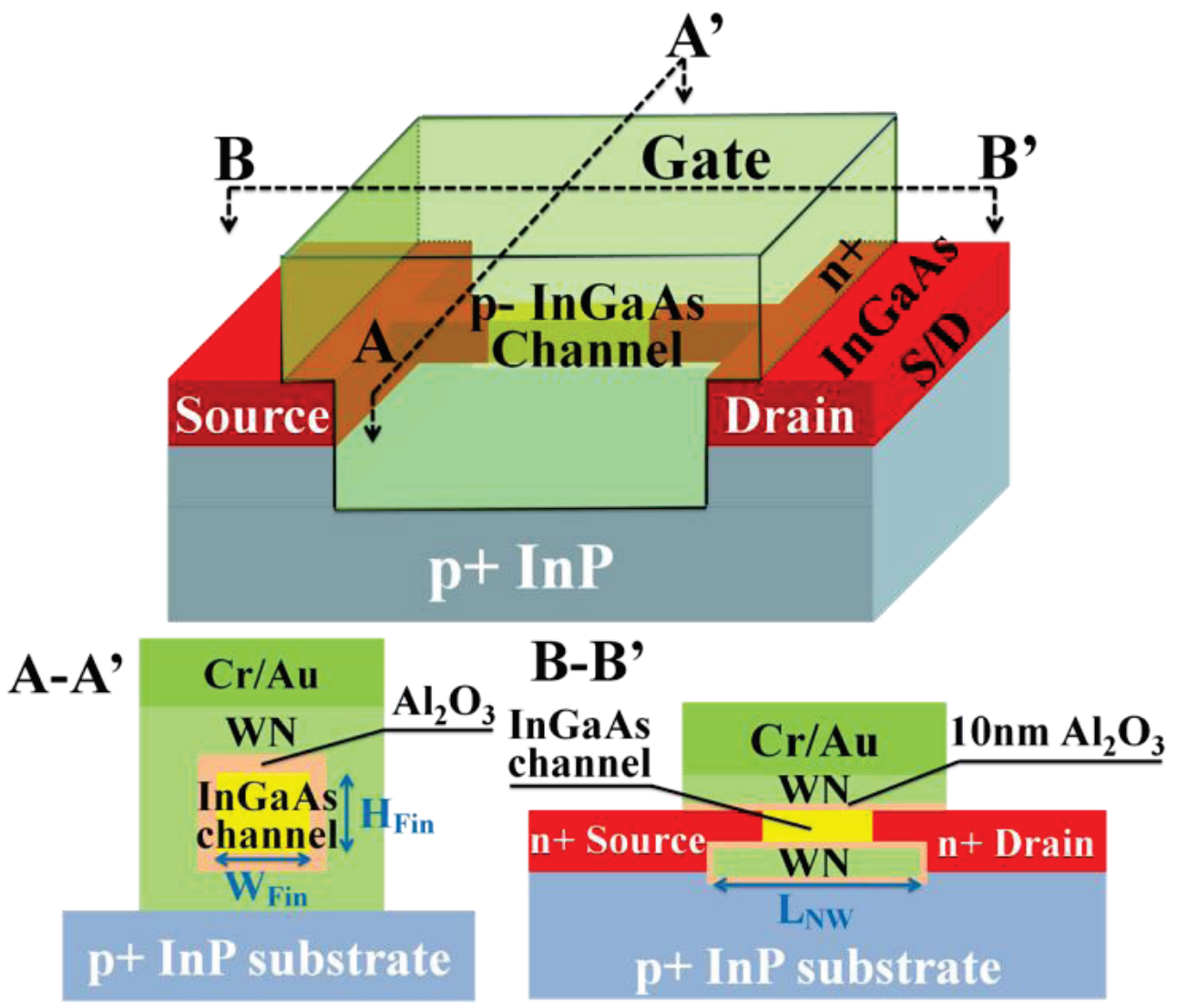

Fig. 1 Schematic view of an inversion-mode GAA n-channel $In_{0.53}Ga_{0.47}As$ ($2\times10^{16}/cm^3$) MOSFET with ALD 10nm $Al_2O_3$/20nm WN gate stacks. A heavily doped wide bandgap InP lies underneath the bottom gate.

- $NH_4OH$ pretreatment and ALD 10nm $Al_2O_3$ as an encapsulation layer
- S/D implantation (Si: 20keV / $1\times10^{14}$ $cm^{-2}$)
- S/D activation (600°C 15s in $N_2$)
- InGaAs fin etching by HDPE $BCl_3$/Ar
- Localized nanowire release by selective etching of InP over InGaAs using HCl based solution
- Pre-gate passivation using $(NH_4)_2S$ solution
- ALD 10nm $Al_2O_3$ and 20nm WN gate stacks conformal deposition surrounding nanowire channel
- Gate pad (Cr/Au) definition and gate etch using HDPE $CF_4$/Ar
- S/D Au/Ge/Ni metallization and 350°C annealing
- Test pad deposition

Table 1 Fabrication process flow for inversion-mode high-k/InGaAs GAA FETs. All patterns were defined by a Vistec VB-6 UHR electron beam lithography system. Dry etching was carried out by a Panasonic E620 high density plasma etcher (HDPE).

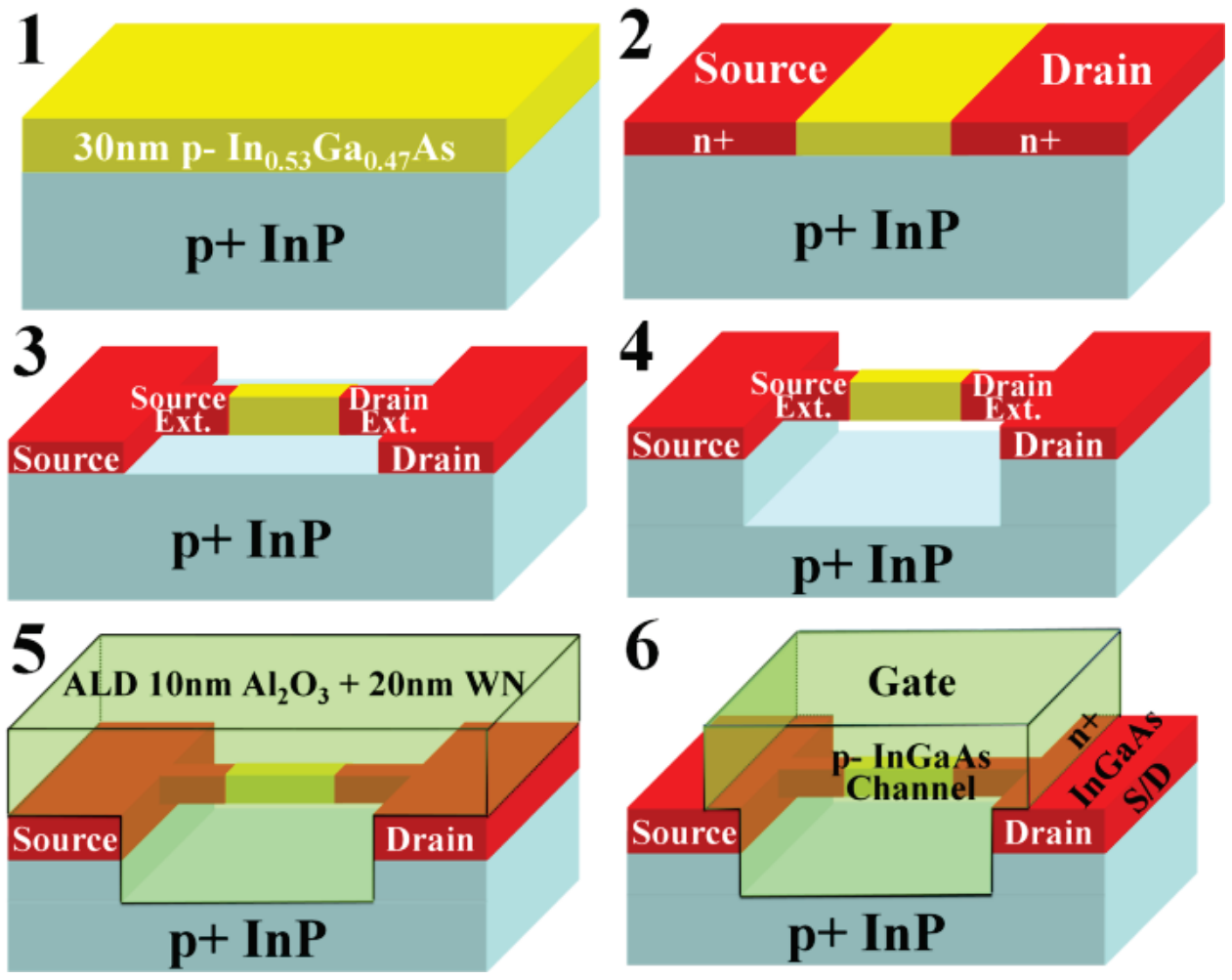

Fig. 2 Schematic diagram of key process steps in the fabrication of InGaAs GAA MOSFETs.

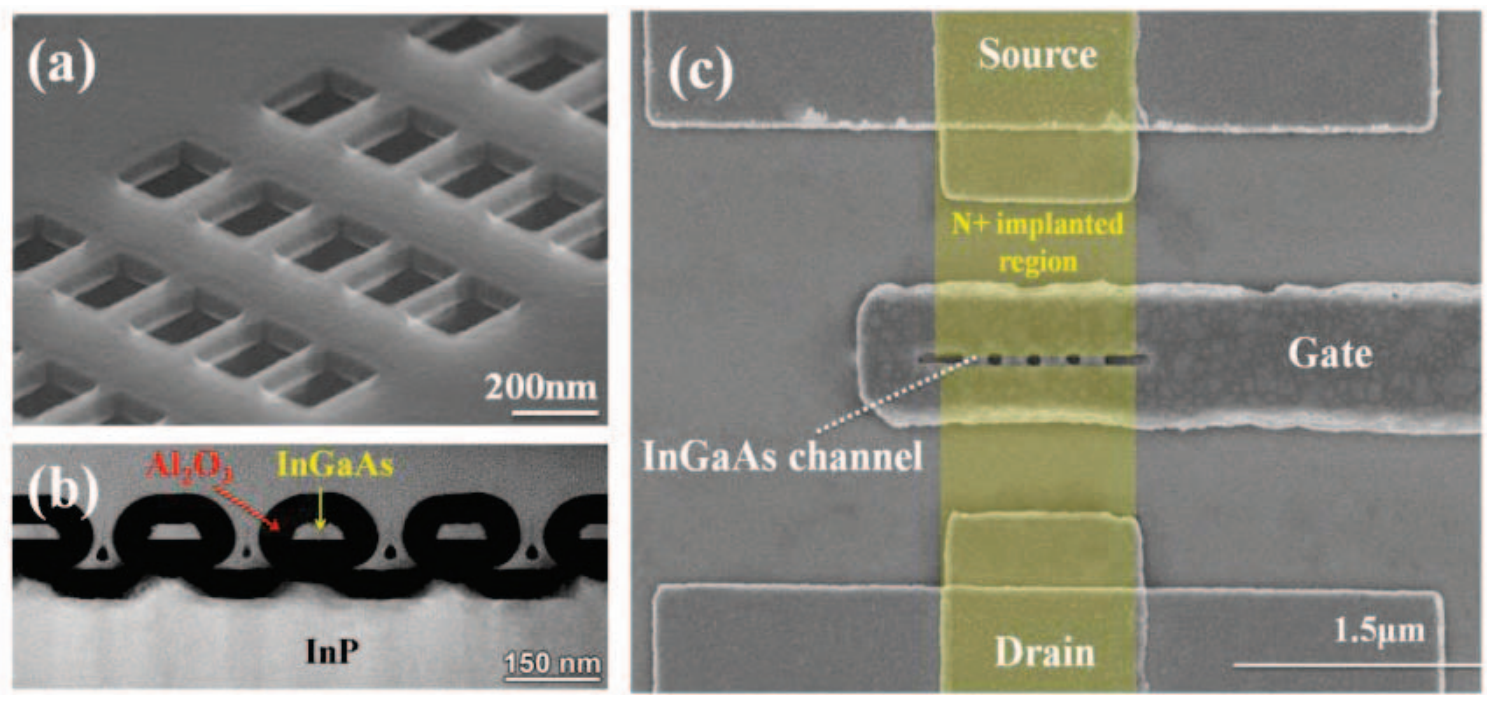

Fig.3 (a) Tilted SEM image of free-standing InGaAs nanowire test structures after the release process. (b) Cross-sectional STEM image of InGaAs nanowire test structures wrapped by 50nm ALD $Al_2O_3$ on InP substrate, confirming the nanowires are completely released (c) Top view SEM image of a finished InGaAs GAA FET with 4 parallel wires of $W_{Fin}$ = 30nm , $L_{NW}$ = 200nm and $L_{ch}$ = 50nm.

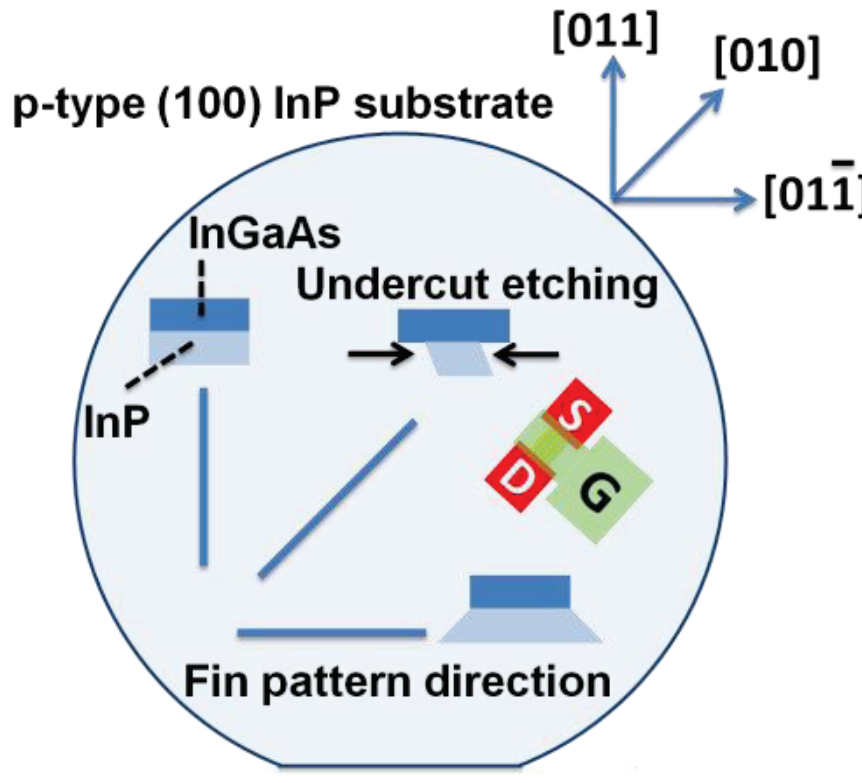

Fig. 4 Schematic diagram of fin patterning direction and device aligment to the substrate.

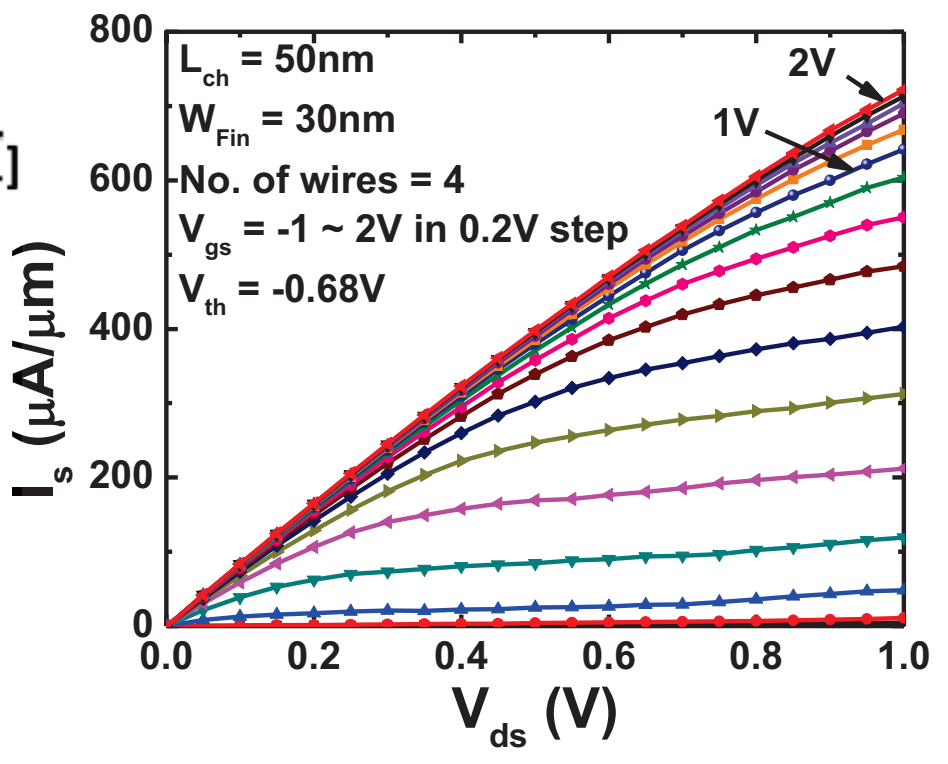

Fig. 5 Output characteristic of a representative GAA FET with $L_{ch}$=50nm, $W_{Fin}$=30nm. The current from each wire is normalized by 120nm, the perimeter of the channel.

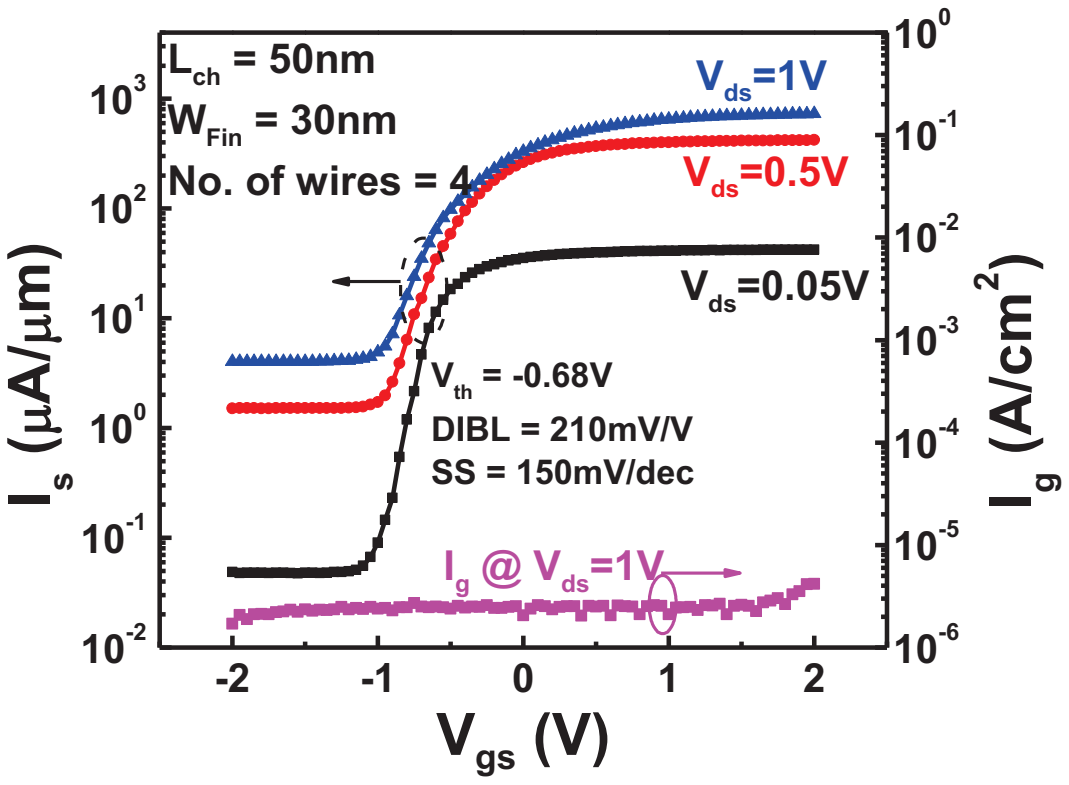

Fig. 6 Transfer characteristic and gate leakage current of a representative GAA FET with $L_{ch}$=50nm and $W_{Fin}$=30nm.

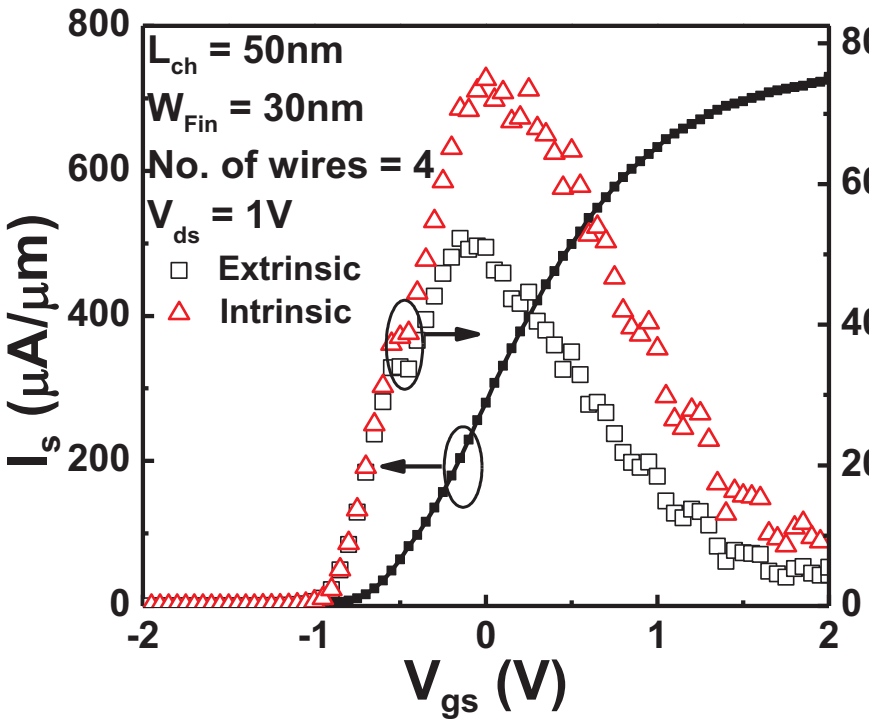

Fig. 7 Current and extrinsic/intrinsic transconductance in saturation regime. The source/drain resistance $R_{SD}$ is extracted to be 1150Ω·μm.

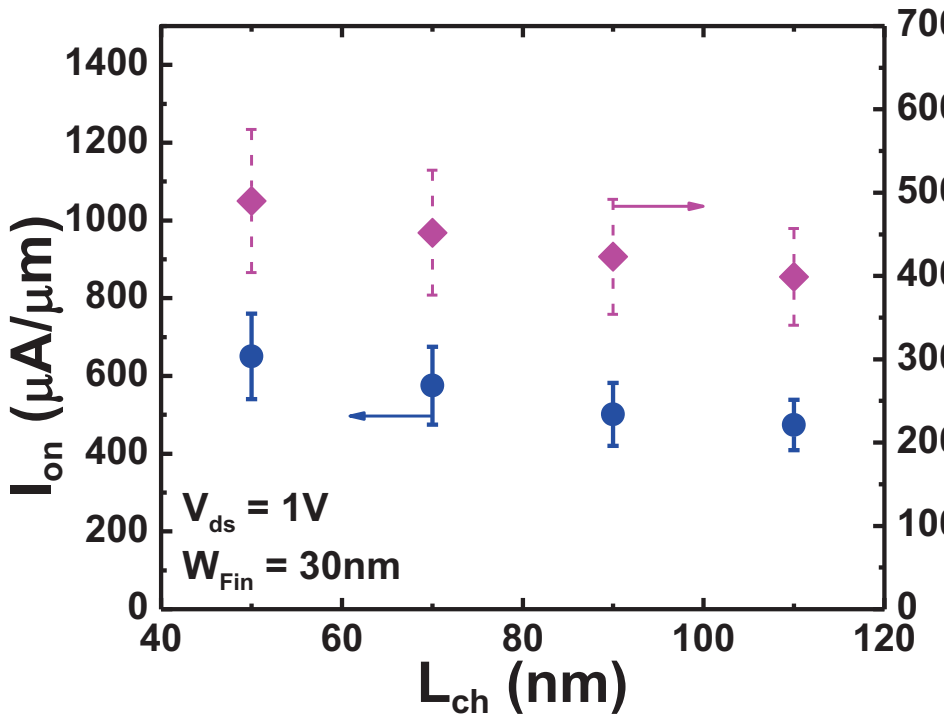

Fig. 8 $I_{ON}$ and $g_m$ vs. $L_{ch}$ for GAA FETs with $W_{Fin}$=30nm. The values are determined by measuring 20 different devices at the same $L_G$. Error bar shows statistical variations of multiple devices.

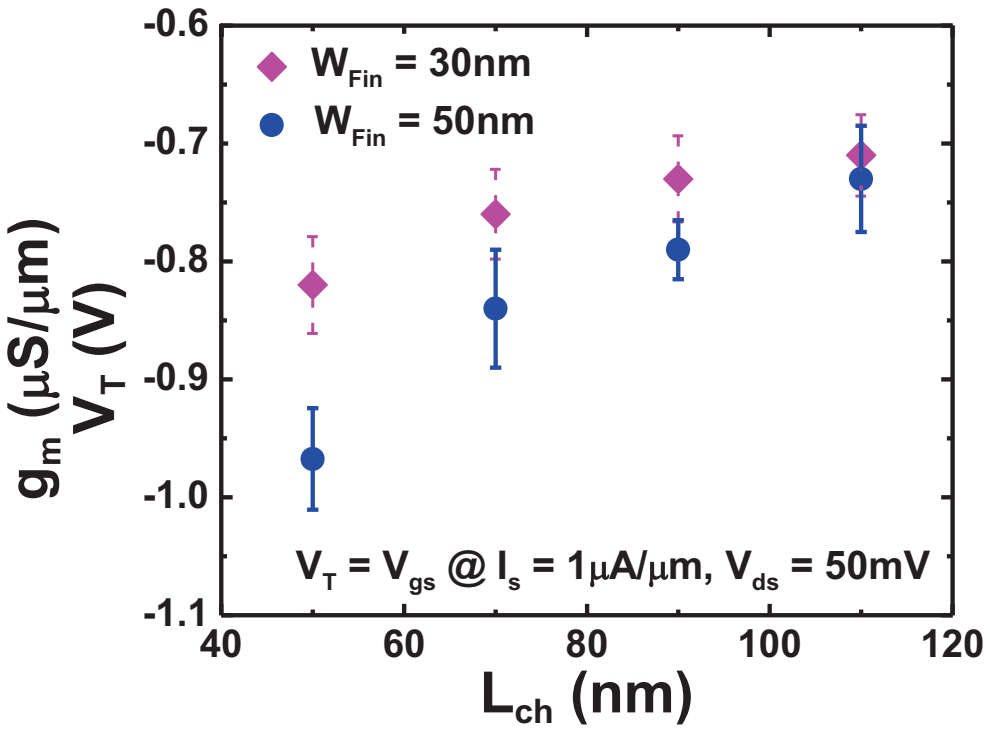

Fig. 9 $V_T$ vs. $L_{ch}$ for GAA FETs with $W_{Fin}$=30nm and 50nm. Smaller $W_{Fin}$ devices show better $V_T$ roll-off.

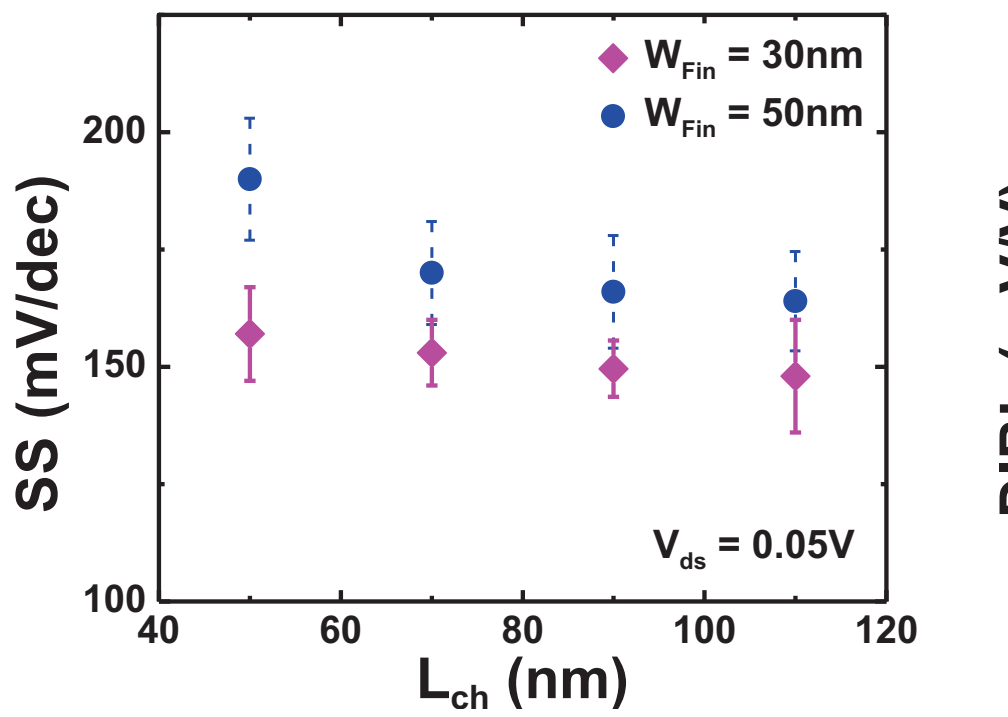

Fig. 10 SS vs. $L_{ch}$ for GAA FETs with $W_{Fin}$=30nm and 50nm. The estimated upper limit of midgap $D_{it}$ from SS is $5.6\times10^{12}$ /cm$^2$·eV, a factor of 2 lower than those reported in [1].

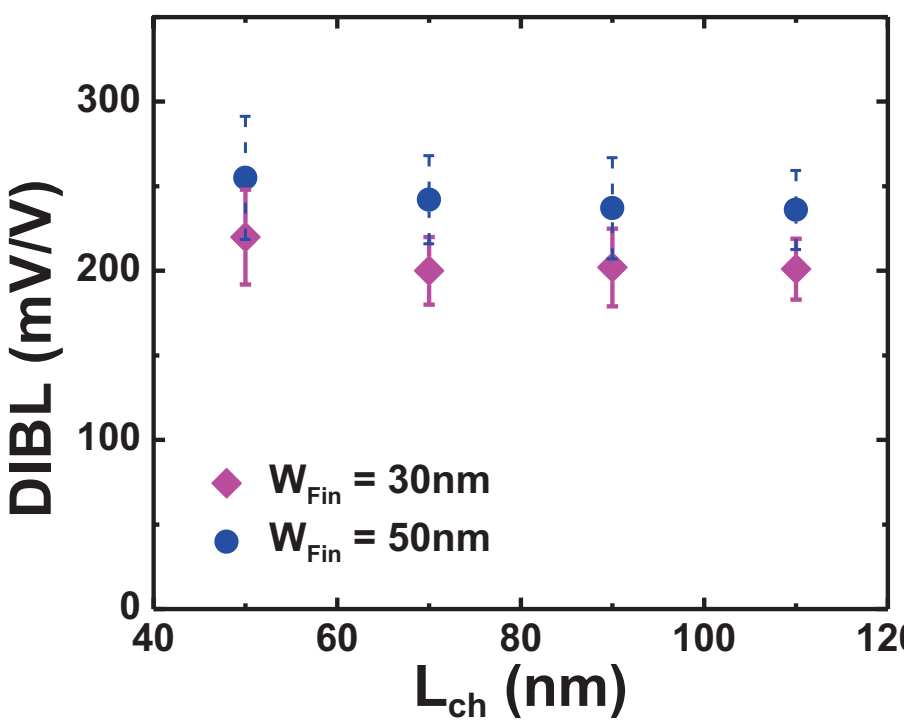

Fig. 11 DIBL vs. $L_{ch}$ for GAA FETs with $W_{Fin}$=30nm and 50nm.

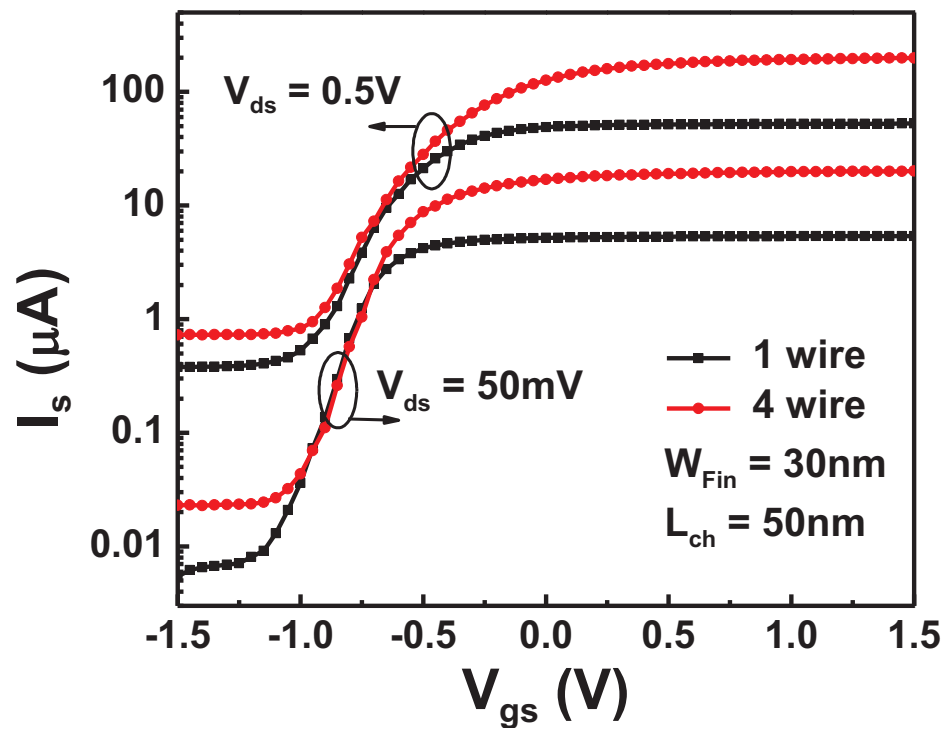

Fig. 12 Transfer characteristic of GAA FETs with 1 and 4 parallel wires.

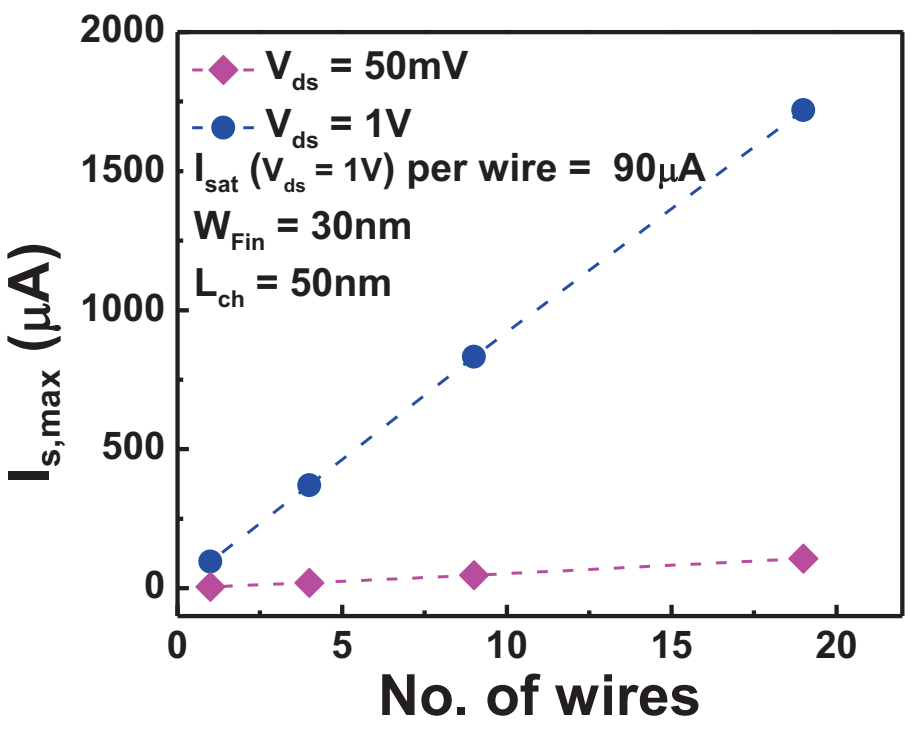

Fig. 13 Linear and saturation current for GAA FETs with different number of parallel wires.

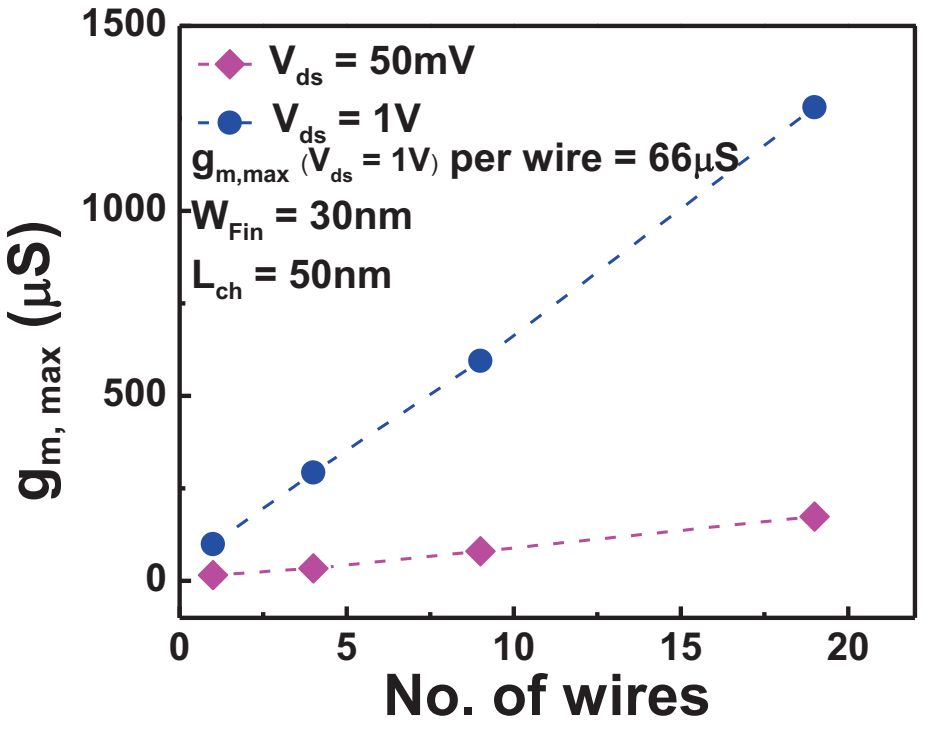

Fig. 14 Linear and saturation transconductance for GAA FETs with different number of parallel wires.

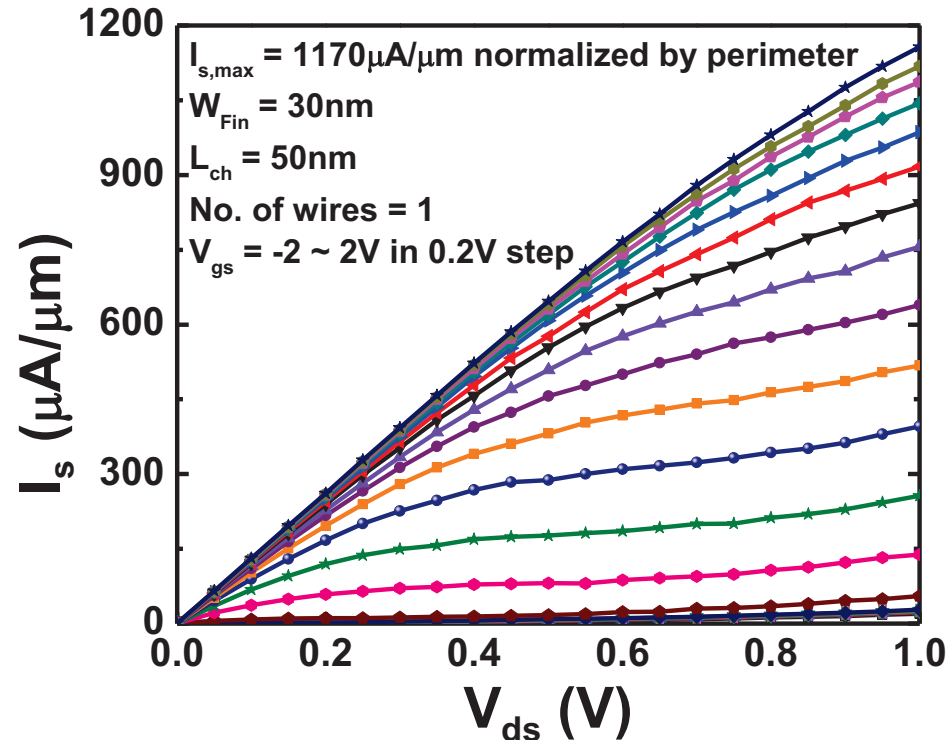

Fig. 15 Output characteristic of a hero GAA FET with saturation current of 1.17mA/μm.

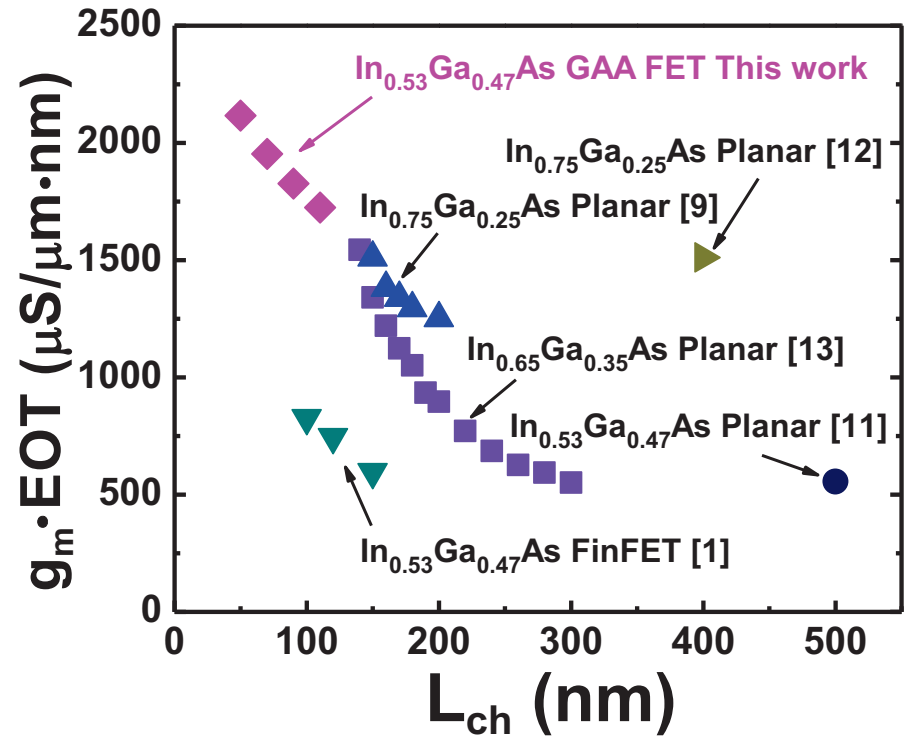

Fig. 16 Benchmarking $g_m$·EOT of planar and non-planar InGaAs surface-channel MOSFETs.

| Structure | $In_{0.53}Ga_{0.47}As$ Planar | $In_{0.53}Ga_{0.47}As$ FinFET | $In_{0.7}Ga_{0.3}As$ FinFET | $In_{0.7}Ga_{0.3}As$ Multi-gate QWFET | $In_{0.53}Ga_{0.47}As$ Gate-All-Around MOSFET |
|---|---|---|---|---|---|
| | Ref [11] | Ref [1] | Ref [2] | Ref [3] | This work |
| Dielectric/ EOT | 8nm ALD $Al_2O_3$ | 5nm ALD $Al_2O_3$ | 19nm MOCVD HfAlO | $TaSiO_x$/1nm InP ($T_{OXE}$ = 2.1nm) | 10nm ALD $Al_2O_3$ |
| $L_{ch}$ [nm] | 500nm | 100nm | 130nm | 70nm | 50nm |
| $W_{fin}$ [nm] | | 40nm | 220nm | 60nm | 30nm |
| $I_{sat}$ [μA/μm] | 430 ($V_{ds}$=2V, $V_{gs}$-$V_T$=3.2V) | 220 ($V_{ds}$=1V, $V_{gs}$-$V_T$=0.85V) | 850 ($V_{ds}$=2V, $V_{gs}$-$V_T$=3V) | ~300 ($V_{ds}$=0.5V, $V_{gs}$-$V_T$≈0.5V) | 720 (1170 Max.) ($V_{ds}$=1V, $V_{gs}$-$V_T$=2.7V) |
| DIBL [mV/V] | 350 | 180 | 135 | 110 | 210 |
| SS [mV/dec] | 240 | 145 | 230 | 120 | 150 |

Table 2 Comparison of InGaAs GAA FETs in this work and recently reported top-down non-planar III-V FETs.